# Quantum Matter in Ultrahigh Magnetic Fields
**A workshop sponsored by the National Science Foundation**

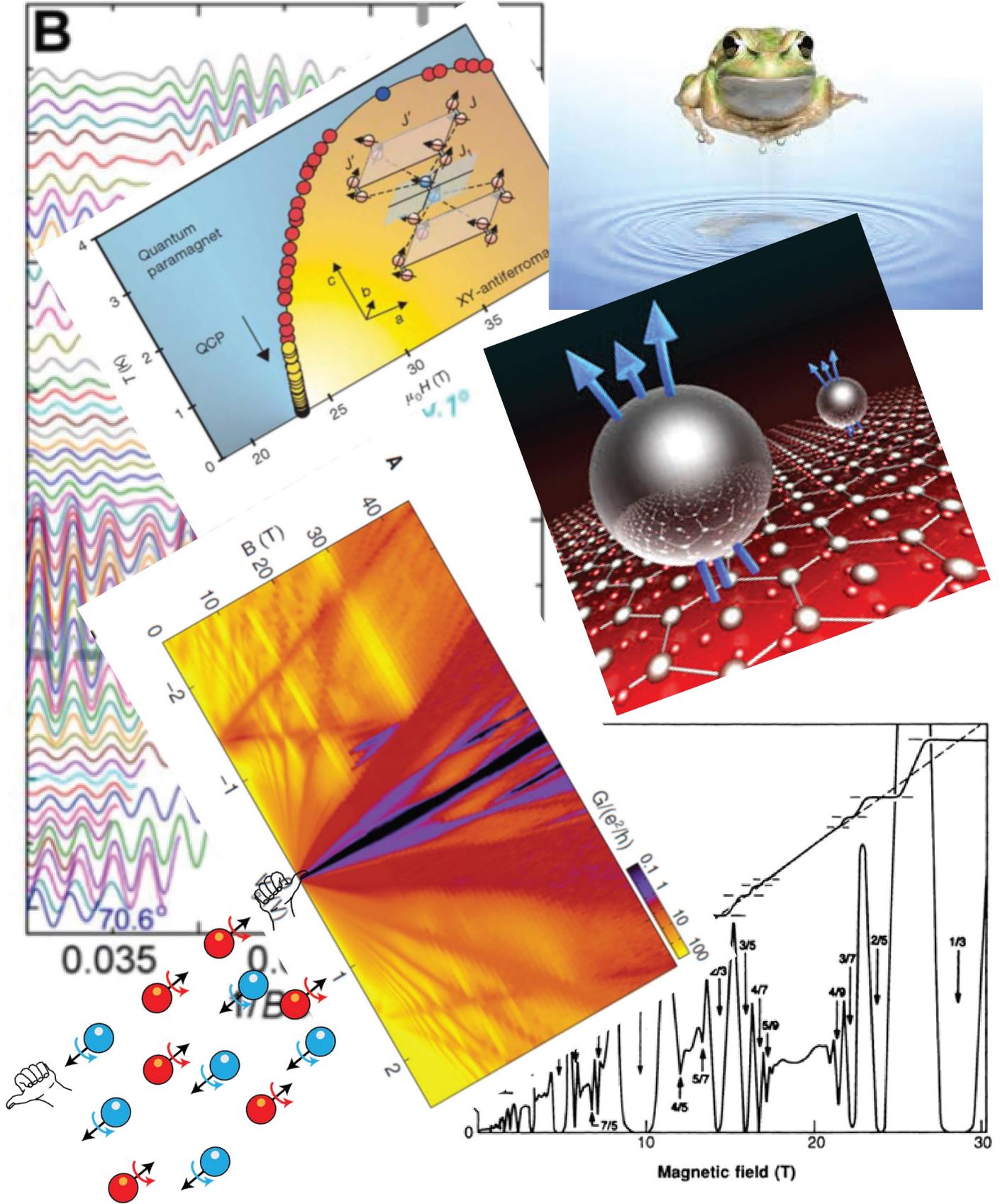

# Quantum matter in ultrahigh magnetic fields
A workshop sponsored by the National Science Foundation

N. P. Ong[1] and Lu Li[2]
Princeton University[1] and University of Michigan[2], Ann Arbor

This material is based on work supported by the National Science Foundation under grant DMR-1745525. The opinions, findings, and conclusions or recommendations expressed in this material are those of the authors, and do not necessarily reflect the views of the National Science Foundation.

*Cover picture credits:*
Levitating frog: ignobel-8-100559187-gallery.idge
Quantum paramagnet: Sebastian *et al*. *Nature* **441**, 617 (2006).
Moire spectrum: Hunt *et al*. *Science* **340**, 1427 (2013).
Red and blue balls: N. P. Ong.
Shubnikov oscillations in cuprate: Sebastian *et al*. *Nature* **511**, 61 (2014).
Fractional Quantum Hall: Störmer, *Rev. Mod. Phys*. **71**, 875 (1999).

**Abstract**

In writing this report we had two goals in mind. The first is to provide a survey of a subset of discoveries from recent experiments performed on quantum matter in high magnetic fields, and to anticipate the scientific opportunities to be realized in even higher fields. Hopefully, the survey will convey a sense of the excitement and pace of high-magnetic-field research in the quantum-matter community to a broader audience (undergraduates, especially). The second goal is to discuss the comparative merits of two options: a pulsed-field facility for attaining a magnetic field of 150 Tesla (of duration 1-10 msec) or a DC field facility that attains 60 Tesla. A workshop involving leading scientists involved with quantum phenomena in high magnetic fields was held at NSF, Alexandria Sep. 21,22 (2017) to address these issues.

We start with a survey of the discoveries in quantum matter. The group of topological matter includes Topological insulators, Topological crystalline insulators (TCI), and Dirac-Weyl semimetals. We discuss the recent experimental observation of the chiral anomaly in Dirac-Weyl semimetals, which is a solid-state analog of a phenomenon first encountered in the decay of π-mesons in high-energy physics. An intense magnetic field will broaden considerably the experimental arena for further exploration of the chiral anomaly. In the TCI rocksalt, PbSnSe, the thermopower is strongly enhanced in strong magnetic fields because of the 3D Dirac electrons. A recent theory predicts that the thermoelectric figure of merit (ZT) can be as large as 10 in high fields. In metals, quantum oscillations in the resistance or magnetization in a strong field has long been the standard technique to measure the caliper of a generic feature of all metals called the Fermi surface. Insulators do not have a Fermi surface. However, recent experiments detecting quantum oscillations in strong magnetic fields in the insulators $SmB_6$ and $YbB_{12}$ pose a challenge to the definition of the insulating state. Graphene has long fascinated scientists because its electrons mimic relativistic (Dirac) electrons moving in a one-atom thick film. High-field experiments on a new generation of high-purity graphene samples have uncovered a number of exotic quantum states. As an example, we describe the Hofstadter butterfly states observed in moire superlattices. Quantum magnets are magnetic materials that are insulating. Nonetheless, a strong magnetic field can tilt these materials into novel quantum states by aligning their electronic magnetic moments. An interesting example is a silicate called "purple Han," which enters, in high field, a quantum state equivalent to superfluidity in liquid helium, but involving the magnetic moments. In certain classes of novel metals, quantum effects cause the electrons to behave like elongated molecules (much like nematic molecules in familiar liquid crystals). In sufficiently strong magnetic fields, these elongated composite objects can also be aligned. Finally, the high temperature cuprate superconductors continue to pose a challenge. We describe how magnetic fields higher than currently available can help resolve many outstanding puzzles.

We close by weighing the relative merits of pulsed versus DC magnetic fields, and discuss the scientific reasons why the choice is not at all clear cut. As a pertinent example, we describe the heavy investments on implosive magnets in the 1970's and their scientific legacy, and discuss how the lessons learned can inform the current debate. The advantages of a third option – building both a pulsed-field facility with lower target field (130 T) and a 60 T DC magnet – are briefly described. Some scientific terms are explained in Appendix A.

# 1. Introduction

The past 2 decades have witnessed the rapid growth of interest and activity in the field of research many researchers call "quantum matter." The growth has been fueled by a spate of experimental and theoretical discoveries, many from unexpected directions. The term quantum matter refers to materials that display dominant quantum effects writ large. The range of phenomena include, for example, many-body states (superconductivity and superfluidity), states stabilized by topological properties and symmetry (edge states, Dirac states, chiral states, Majorana bound states, Haldane gap systems), quantum states with strong fluctuations arising from competing phases, states in which the electrons are replaced by emergent particles (composite fermions and bosons), magnetic states describable by the condensation of hardcore bosons, and highly frustrated magnets that may harbor the long-sought quantum spin liquid. At a superficial level, some aspects of these phenomena and the tools employed have been researched (or used) for decades on familiar materials. However, the distinction is not merely putting lipstick on a pig. In the emerging field of quantum matter, practitioners share a comprehensive theoretical overview that emphasizes the exploration of inherent quantum effects. The modern perspective is sharply distinct from the older one; experiments on the fractional quantum Hall effect (FQH) in GaAs-AlGaAs quantum wells are informed by a radically different theoretical perspective than that used to analyze, say, high-field transport experiments on GaAs in the 1960s.

In tracing the roots of developments and discoveries in this area, we may identify the two major rivers of research topics -- the Integer and Fractional Quantum Hall effects and cuprate superconductivity -- that dominated condensed matter physics throughout the 1990s and the early 2000's. The circle of new ideas spawned by these two fields were further enriched with the discovery of graphene in the mid 90's. A decade later, infusion of concepts from differential topology and quantum field theory in the field of topological quantum matter (topological insulators, topological crystalline insulators, Dirac/Weyl semimetals and nodal semimetals) has further accelerated the pace of research and discovery. In hindsight, the continuity of ideas and their gradual evolution from the Quantum Hall era to the more current topics is quite apparent (the Appendix explains some of these topics). In the majority of these discoveries, strong magnetic fields have either played an intrinsic role in the discovery process, or have been essential for exploring further their consequences.

# 2. Large-ZT thermoelectrics in PbSnSe in the quantum limit

The rocksalt PbSnSe belongs to the group of materials called topological crystalline insulators [1-5]. In this group, crystalline symmetry (e.g. mirror symmetry) protects surface Dirac states if their nodes occur at the mirror planes. In addition to the protected surface states, electrons occupy bulk states (at the $L$ points of the Brillouin zone), which behave like 3D Dirac fermions with a small effective mass. A recent theory predicts that the bulk electrons in the quantum limit can exhibit record high thermoelectric figure of merit $ZT$.

For thermoelectric cooling, the dimensionless parameter $ZT = S^2\sigma T/\kappa$ sets the maximum temperature difference achievable for a given thermoelectric material ($S$, $\sigma$ and $\kappa$ are, respectively, the Seebeck coefficient, electrical conductivity and thermal conductivity of the material). To date, the best thermoelectric materials have $ZT \sim 2.4$ at elevated temperatures $T$ (>400 K). Skinner and Fu [6] recently proposed that Dirac semimetals in the extreme quantum limit can exhibit a thermopower that increases without limit in intense fields. The Seebeck coefficient is given by the ratio $S_{xx} = J^Q/TJ^e$, with $J^Q$ and $J^e$ the heat and charge current densities, respectively. In the quantum limit (in a transverse magnetic field **B**), the energy currents are additive for holes and electrons whereas the charge currents subtract. As a consequence, the ratio $J^Q/J^e$ increases monotonically with $B$ without saturation.

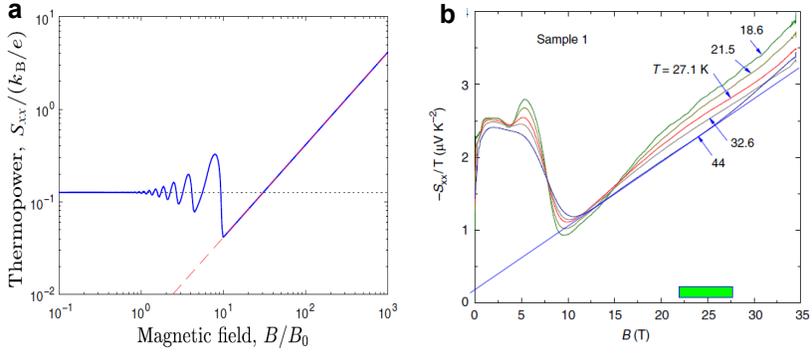

Figure 1: (a) Seebeck coefficient $S_{xx}$ vs. $B$ calculated for a Dirac/Weyl semimetal in the quantum limit. Oscillations reflect emptying of Landau levels (Skinner and Fu [6]). (b) The measured $S_{xx}$ vs. $B$ in the rocksalt PbSnSe up to 35 T, from 19 to 44 K (Liang et al. [5]).

For a Dirac/Weyl semimetal, Skinner and Fu [6] calculate that $S_{xx}$ increases linearly with $B$ (Fig. 1a). They point out that the linear increase is already evident in measurements on PbSnSe by Liang et al [5] at $T$ as high as 44 K (Fig. 1b). In still higher magnetic fields, $ZT$ values as large as 10 are expected. The availability of next-generation magnetic fields will open a new frontier on large-$ZT$ research in semimetals featuring 3D (massive) Dirac electrons in the quantum limit.

### 3. Quantum oscillations in correlated insulators

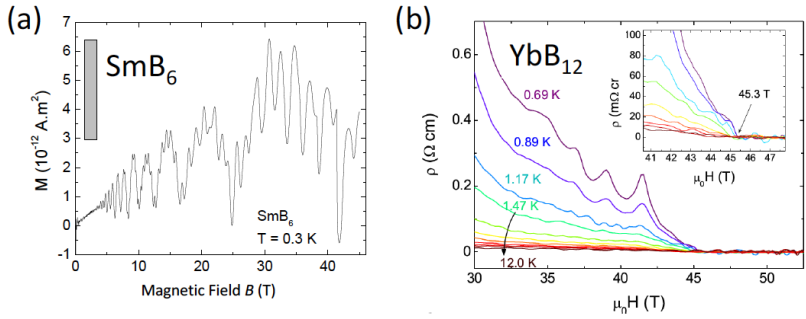

Figure 2: Quantum oscillations in correlated Kondo insulators. (a). Oscillations observed in torque magnetization (Xiang et al. [9]). (b). Pulsed field measurements of the resistivity of $YbB_{12}$ showing quantum oscillations starting near 30 T. At 45.3 T, the field induces the energy gap to close (Xiang et al. [10]).

When is an insulator an insulator? Recent experiments in very high magnetic fields are challenging long-held notions of the insulating state [7-10]. In a magnetic field, the free electrons in a metal occupy a series of bands called Landau levels (Appendix). In a steadily increasing $B$, the successive emptying of the higher Landau levels produces quantum oscillations in the observed resistance or magnetization. These oscillations have long provided a direct measurement of the "caliper" diameter of a metal's Fermi surface. They also reveal when the chemical potential enters the lowest Landau level (the quantum limit). Quantum oscillations in an insulator (which does not have free electrons or a Fermi surface) would seem to be oxymoronic. However, a series of experiments on the correlated "Kondo" insulators $SmB_6$ [7-9] and $YbB_{12}$ [10] have uncovered pronounced oscillations at low temperatures. As the temperature $T$ decreases, hybridization between localized f bands and itinerant d bands, together with interaction effects, opens a narrow gap, resulting in a thermally activated resistivity. In both materials, the abrupt saturation of the resistivity below 5 K suggests the co-existence of metallic surface states and insulating bulk states. Quantum oscillations were initially detected by magnetization in $SmB_6$ and attributed to the surface states [7], but hints of a higher frequency component were later detected and associated with the insulating bulk states [8]. The debate of surface *vs* bulk origin was informed by subsequent detailed measurements of the angular dependence of the oscillations in "shaped" crystals taken in fields up to 45 Tesla [9] (Fig. 2a).

The mystery has deepened with new results obtained on YbB$_{12}$. Experiments [10] performed in DC fields of 45 T and pulsed fields to 65 T (Fig. 2b) have uncovered quantum oscillations in both the resistivity and magnetization (the two sets of oscillations seem to display distinct symmetry properties as the field **B** is tilted). The results suggest an exotic quantum state in which one set of oscillations arises from a putative bulk insulating state. Access to DC magnetic fields above 50 T and pulsed magnetic fields above 100 T will help uncover further exotic results in correlated insulators that challenge the text-book definition of the insulating state.

4.  **Topological matter in high magnetic field**

The Dirac/Weyl (DW) semimetals are a novel class of quantum materials in which the electrons mimic the linear energy-momentum dispersion of relativistic ("massless") electrons, as found for the Dirac electrons in graphene and topological insulators (TIs). Unlike the latter 2D materials, the fermions in DW semimetals are 3D. The increase in dimensionality introduces a host of new physics long investigated in relativistic quantum field theory. The breaking of time-reversal invariance in a magnetic field splits each Dirac cone into two Weyl cones, which have opposite handedness (or chirality). This occurs in the semimetals Na$_3$Bi, [11] Cd$_3$As$_2$ and ZrTe$_5$ [12,13] as well as in the half-Heusler GdPtBi [14]. In a second group of "Weyl semimetals" (TaAs, NbAs, TaP, NbP), the Weyl nodes are already present in zero field because inversion symmetry is broken in the lattice [15-17]. In either case, a novel property of Weyl fermions emerges in the quantum limit reached only by the application of a strong magnetic field.

Unlike the higher Landau levels, the lowest Landau level ($N$=0) of Weyl fermions is invariably "chiral" (it disperses in only one direction). The velocity is strictly parallel or antiparallel to **B** if the Weyl fermion is right-handed or left-handed, respectively. Hence, in the quantum limit, the Weyl electrons segregate into 1D right- and left-handed fermions that cannot reverse their respective velocities (Fig. 3a). One of the new results found is the "chiral anomaly," which expresses the appearance of a novel "axial" current when an electric field **E**, applied parallel to **B**, leads to a strong imbalance between left- and right-moving populations.

The chiral anomaly first appeared in the theory of pion decay (Adler, Bell and Jackiw) which explained why neutral pions $\pi^0$ decay 300 million times faster than the charged pions $\pi^{\pm}$.

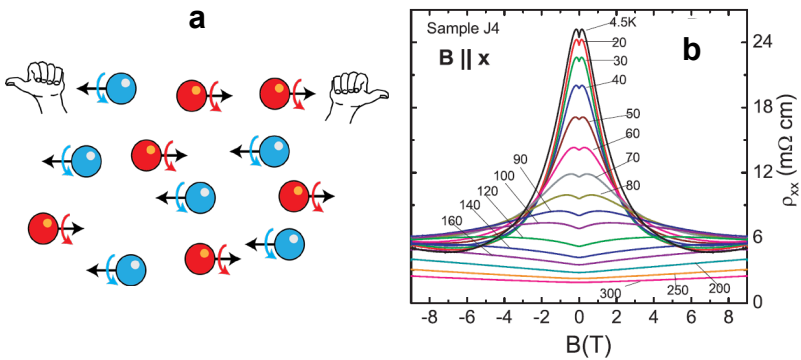

Figure 3: (a) In the quantum limit, Weyl fermions segregate into left- and right-handed populations (blue and red spheres, respectively) with opposite velocities (black arrows). (b) The chiral anomaly observed in Na$_3$Bi as a decrease in the resistivity $\rho_{xx}$ at selected temperatures with **E**||**B**. Xiong *et al*. [11]). Higher fields will access a wider range of Weyl systems that exhibit other physics due to the chiral anomaly.

The vast difference exists because the chiral anomaly creates the rapid decay channel, $\pi^0 \rightarrow \gamma\gamma$ ($\gamma$ are photons), by coupling the pion field to the electromagnetic field ($\pi^{\pm}$ cannot access this channel because of charge conservation). In 1983, Nielsen and Ninomiya predicted that the chiral anomaly should appear when 3D Dirac states are realized in crystals.

In the semimetals, Na$_3$Bi, GdPtBi and ZrTe$_5$, the quantum limit is reached in fields $B$ < 8 Tesla because the carrier populations are small. The resulting axial current (the analog of $\pi^0 \rightarrow \gamma\gamma$) leads to a steep decrease in the resistivity $\rho_{xx}$, observed [11] as a negative longitudinal magnetoresistance LMR (Fig. 3b). The interpretation of the negative LMR as the chiral anomaly is supported by recent experimental tests [18].
Magnetoresistance experiments in the Weyl semimetals TaAs and NbP are seriously hampered by their larger carrier populations and an interfering classical effect called "current jetting" [18] which leads to a strongly inhomogeneous current-flow pattern (especially pronounced in high-mobility samples). However, the chiral anomaly has signatures beyond magnetoresistance, for example, in measurements of the magnetoplasmon dispersion [19], thermoelectric response, electronic thermal conductivity and heat capacity. An interesting proposal is that a Weyl crystal can be cooled by a process dubbed "dechiralization," which is analogous to the nuclear demagnetization process for cooling a magnetic salt [20]. DC and pulsed magnetic fields in the 50 T to 150 Tesla range will greatly facilitate these experiments, and open a new chapter in the experimental investigation of chiral anomaly signatures in Weyl semimetals.

5. **The Hofstadter Butterfly and Fractional States in Graphene**
Graphene, comprised of a single layer of carbon atoms in a hexagonal lattice, has achieved iconic status in the popular media because it is promising for many applications despite its structural simplicity. From the physics viewpoint, the Dirac states in graphene have been of intense fundamental interest from the start [21,22], in part because the states display a linear energy-momentum dispersion just like relativistic fermions. In the past decade, this interest has only grown, stimulated by the observation of several exotic quantum states in magnetic fields of 10-40 Tesla. The most recent advances were achieved on ultra-clean graphene films sandwiched between hexagonal boron nitride (hBN) crystals.

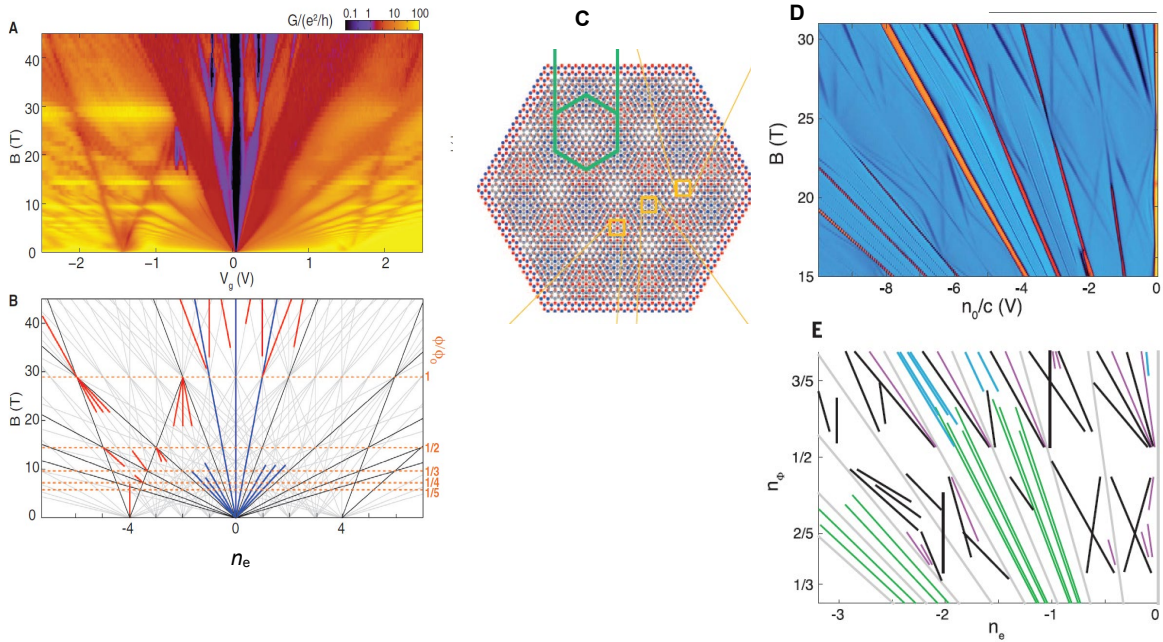

Figure 4: (A) The linear trajectories of mini-gaps in graphene-hBN superlattice in the $n$-$B$ plane. (B) The equivalent skeletal plot in the $n_e$-$n_\phi$ plane (Hunt et al. [25]). (C) The moire pattern formed by superposing graphene (grey circles) on boron nitride (red and blue circles). The green hexagon is the superlattice unit cell. (D) The gap trajectories obtained by capacitance measurements in $n_e$-$B$ plane (Spanton et al. [26]) In Panel (E) the trajectories are replotted in the $n_e$-$n_\phi$ plane. Black lines show new fractional states with $s = ½$ converging to points on the line $n_\phi = 1/2$, and also 2/5. Green lines represent FQH states. Blue lines are novel fractional Chern Insulator states.

When exfoliated graphene is sandwiched between two boron nitride crystals, the slight (1.8%) lattice mismatch creates a "beating spatial frequency" that appears as a moire pattern with a superlattice period λ = 7-11 nm (Fig. 4C). The moire pattern provides a unique opportunity to investigate the quantum behavior of electrons moving in a superlattice periodic potential subject to an intense magnetic field. In the quantum regime, the applied magnetic field $B$ is expressed as the flux density $n_\phi$ (the number of flux quanta piercing a unit area). In the absence of the periodic potential, each Landau level is a flat band with negligible broadening.

The moire potential splits the level into a series of mini-bands separated by a hierarchy of self-similar mini-gaps [23-25]. Hofstadter predicted that, with increasing magnetic field (or $n_\phi$), the mini-gaps obey the linear relation $n_e = tn_\phi + s$, where $n_e$ is the number of electrons occupying a superlattice unit cell, and $t$ and $s$ are integers. The energies of the mini-bands plotted vs. $B$ reveals a rich tapestry of fractal states dubbed the Hofstadter butterfly. Experiments [23-25] performed on graphene-hBN superlattices in fields up to 45 Tesla have confirmed the original Hofstadter predictions, and also uncovered a new regime where $t$ and $s$ can assume fractional values periodic potential

The experimental results of Hunt *et al.* [25] (Fig. 4A) are compared with the Hofstadter butterfly predictions in Fig. 4B. Plotting $B$ (or $n_\phi$) vs $n_e$, we see that the mini-gaps follow linear trajectories. The fan of black lines converging to the origin $(n_e, n_\phi) = (0,0)$ represent the gap behavior in the absence of the periodic potential with $s=0$ and $t = 0, \pm 1, \pm 2, ...$). These are the conventional IQH states. Blue lines with fractional $t$ are the FQH states. The more interesting set of lines converging at $(n_e, n_\phi) = (\pm 4, 0)$ represent the Hofstadter states with $t = \pm 2$ and $s = \pm 4$. States with other integer values of $s$ are also resolved in Fig. 4A (red lines).

Further improvements of the electron mobility in graphene hBN superlattices have allowed finer details of the minigap spectrum to emerge in fields up to 35 Tesla [26, 27]. Figure 4D shows the trajectories inferred from capacitance measurements (Spanton *et al.* [26]). The skeletal plots of the trajectories in the $n_e$-$n_\phi$ plane (Fig. 4E) uncover states in which $t$ and $s$ assume fractional values. The black lines converging to points on the line $n_\phi = ½$ all have $s = ½$. These are Chern insulator (CI) states (Appendix) specifically created by the moire potential (analogs of the filled Landau band in the QHE problem). In addition, the experiment reveals *fractional* Chern insulator states (FCI, blue lines). These are the analogs of composite particles in the FQH problem. As in the uniform FQH problem, the FCI states are stabilized by strong Coulomb interaction which can stabilize new quantum states in which the charge density develops a spatial modulation with a period longer than the superlattice period.

The rich array of quantum states reflects the elaborate behavior of interacting electrons that attempt to satisfy the conflicting demands of the moire potential and the field quantization. By going to higher $B$ (larger $n_\phi$), we anticipate more exotic states to emerge.

## 6. Triplet Condensation in Quantum Magnets

In a strong magnetic field, the Zeeman energy of spins in a lattice can become large enough to force the sample into novel many-body states. An elegant example is provided by torque experiments by Sebastian *et al.* [28] on the quantum magnet, barium copper silicate $BaCuSi_2O_6$, the main ingredient of "Han Purple" (a vivid dye restricted to royalty in ancient China).

$BaCuSi_2O_6$ is comprised of layers of $Cu^{2+}$ pairs or "dumb-bells" (Fig. 5, insert) arrayed in a square lattice (in the *a-b* plane). The spins in each dumb-bell form a singlet pair. Throughout the blue region, the singlet pairs are disordered because of incompatibility between long-ranged spin orderings in adjacent layers (geometric frustration). At finite *T*, the spins can be excited to the triplet states, which are separated from the singlet state by an energy gap. As *B* increases, the Zeeman energy lowers one of the triplet states towards the singlet (Fig. 6a), eventually closing the gap at the phase boundary in Fig. 5. Classically, one then anticipates an inhomogeneous mixture of sites in the triplet state and others in the singlet state. However, in the limit $T \rightarrow 0$,

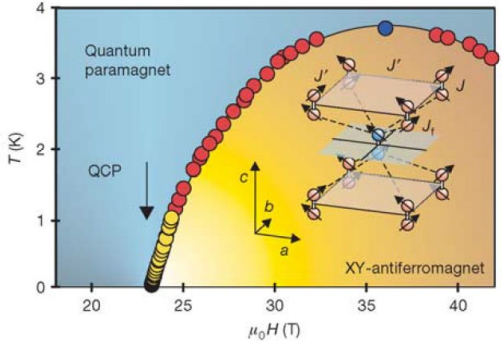

Figure 5: The phase boundary (yellow and red circles) between the paramagnet and antiferromagnet phases in BaCuSi$_2$O$_6$. The inset shows spins of Cu$^{2+}$ ions forming "dumb-bell" singlets. In the blue region, ordering of the singlets is precluded by geometric frustration. In the yellow region, condensation of triplet excitations leads to 2D canted antiferromagnet. (Sebastian et al. [28])

quantum effects dominate. A powerful mapping approach is to regard each triplet site as occupied by a hardcore boson and the singlet sites as unoccupied. As $T \rightarrow 0$, the bosons undergo Bose-Einstein condensation to a long-range ordered state (yellow region). For the original spin problem, this implies that every site has the *identical* coherent linear combination of triplet and singlet states (analogous to superfluidity in liquid helium). The authors find that the effective dimensionality decreases from 3 to 2 which they dub dimension reduction at a quantum critical point [28].

In a later experiment, Haravifard *et al*. [29] have observed a cascade of sharp transitions followed by magnetization plateaus in the dimerized quantum magnet SrCu$_2$(BO$_3$)$_2$, in fields up to 40 Tesla. The authors interpret the pattern as the emergence of fractionally filled bosonic crystals in a mesoscopic superlattice pattern. Recent calculations reveal that the magnon bands in SrCu$_2$(BO$_3$)$_2$ feature a flat band aligned with the node of a Dirac band (Fig. 6b). The appearance of Berry curvature physics has led to predictions [30] of a large thermal Hall conductivity (Fig. 6c).

Clearly, the paradigm of mapping quantum spins to hardcore bosons is a very promising approach that illuminates quantum magnetism and boson physics, especially when we include topological concepts. These systems are experimentally unique because the magnetic field reversibly tunes the number of hard-core bosons to any desired value. DC and pulsed magnetic fields to 50 T to 100 T, respectively, will expand the list of quantum magnets that exhibit similar or related Bose condensation phenomena.

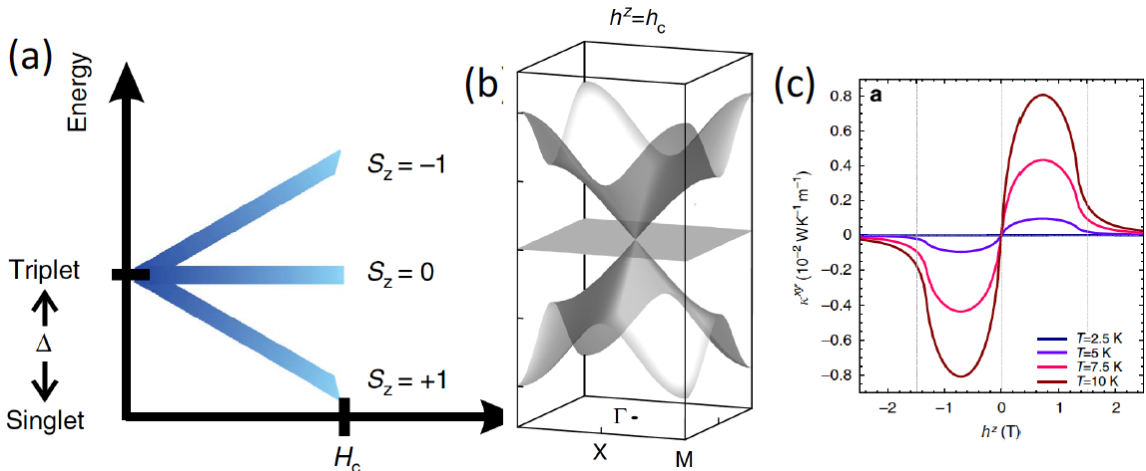

Figure 6. Magnetic field driven triplet condensation in quantum magnets. (a) The effect of a magnetic field on the triplet states. At the critical field $H_c$, the energy of the $S_z = 1$ component falls below the singlet state triggering the condensation of triplets. (b) The calculated magnon bands in SrCu$_2$(BO$_3$)$_2$ [30] reveals band-touching at $H_c$. (c) The band-touching is predicted to produce a large thermal Hall effect [30].

## 7.     Field Enhancement of the Electronic Nematic.

Liquid crystals underpin one of the major consumer display technologies. In the presence of an applied electric field **E**, the long-chained nematic molecules align to create a large anisotropy in the optical reflectivity. The demonstration of similar "liquid-crystal" behavior in electrons in quantum matter is currently a very active area of investigation [31,32]. In the presence of strong interactions, electrons can form composite objects that act as extended objects called nematogens, which can be aligned with **E**. Researchers are actively investigating a causal relation between electronic nematicity and high-temperature superconductivity. Analytis and Fisher [31] demonstrated that nematic fluctuations diverge as $Ba(Fe,Co)_2As_2$ is doped towards the quantum critical point and the nematicity scales with $T_c$ upon doping. In a collaboration between Los Alamos National Laboratory, the National High Magnetic Field Laboratory and the Max-Planck-Institute for Chemical Physics of Solids in Germany, Ronning and Moll [32] used magnetic fields up to 65 T to induce an electronic nematic state in the heavy fermion superconductor $CeRhIn_5$ (Fig. 7). This offers the exciting possibility to study directly the interplay between superconductivity and nematicity, as the nematic state can be directly controlled and tuned by the magnetic field. These recent advances uncover a new aspect of the strongly correlated electron problem and electronic nematicity may lead to new applications in quantum information technology, much as classical nematicity has achieved in display technology.

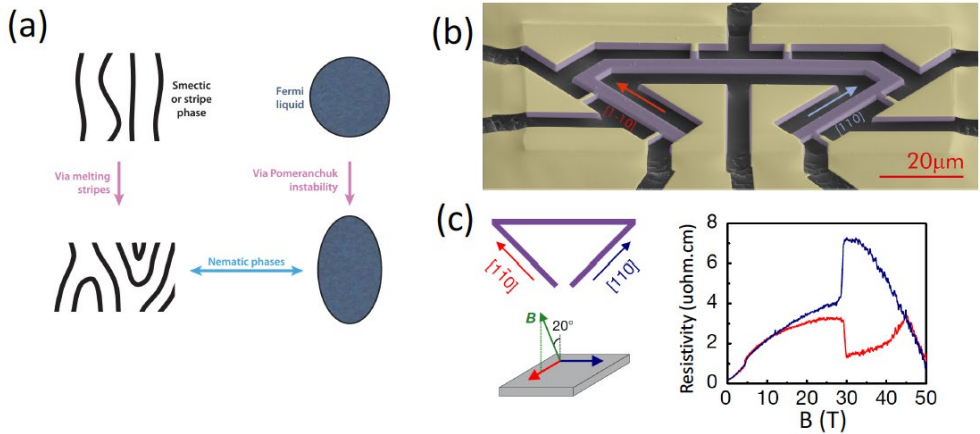

Figure 7. Electronic nematicity enhanced by ultrahigh magnetic fields. (a). A smectic or stripe order may lead to an electronic nematic phase. (b). A sketch of a nanostructure of $CeRhIn_5$ fabricated with Focus Ion Beam etching. (c). With the high magnetic field in different orientations, the magnetoresistance reveals a huge difference of the resistance between the (1 -1 0) and (1 1 0) directions (Ronning et al. [32]).

8. **Cuprate physics in intense fields**

Cuprate superconductivity continues to pose major challenges to researchers despite more than 30 years of intense research. Ultrahigh magnetic fields $H$ (100 T to 150 T) will create experimental opportunities to resolve some of the outstanding issues.

The central puzzle is the pseudogap phase [33] in the underdoped regime where strong competition exists between *d*-wave superconductivity and a host of proximal electronic states (nematic phase, loop-current state, charge density wave, and pair density wave) [34]. The competition engenders strong fluctuations in virtually all observables, which are exacerbated by a fundamental feature of the pair condensate in cuprates: the pair binding energy is extremely large (40 meV or more) but its phase rigidity is much weaker than in low-$T_c$ superconductors (because of the low carrier density). Hence the pairing is extremely robust against unbinding in an applied magnetic field $H$, but the superfluid coherence is less robust because of weak phase rigidity.

From Nernst [35] and torque magnetization experiments [36], researchers have inferred that the pair condensate amplitude survives to temperatures well above the critical temperature $T_c$ (up to 180 K in the underdoped regime). The implication is that, at $T_c$, the pair condensate loses phase rigidity (the property essential for Meissner screening) but – surprisingly - continues to exist 60 to 80 K above $T_c$. Tellingly, the magnetization curve reveals that the diamagnetic response smoothly extends above $T_c$, qualitatively different from mean-field behavior and the well-known behavior of the low-temperature superconductors.

This scenario first discovered in high magnetic fields has been corroborated by a growing list of zero-field experiments: scanning tunneling microscopy (STM) [37], far-infrared reflectivity experiments [38], tracking the 40-meV resonance by neutron spectroscopy [39], and angle-resolved photoemission spectroscopy [40].

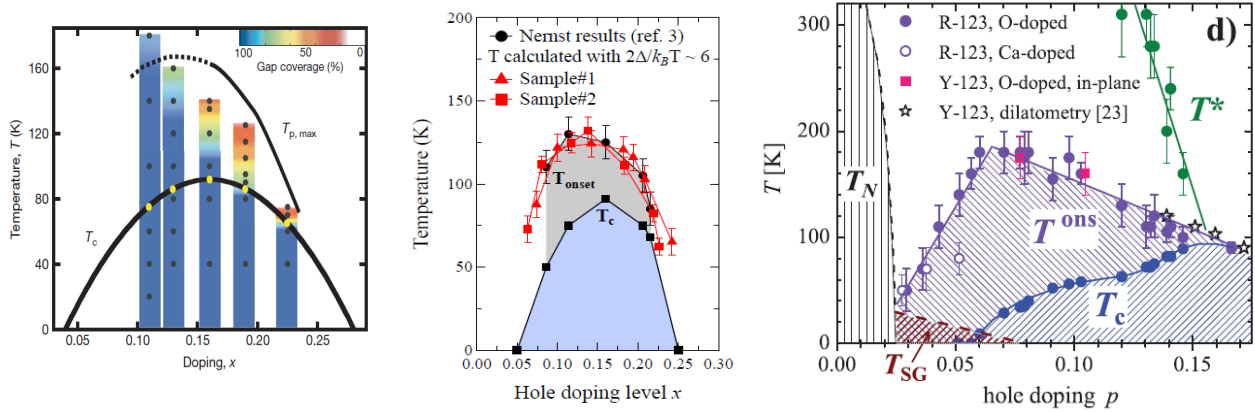

Figure 8: Spectroscopic evidence of pair formation above $T_c$ dome in bilayer cuprates. Panel (a) shows the phase diagram of Bi 2212 in the x-T plane of pair formation observed by STM experiments (Gomes *et al*. [37]). Panel (b) shows the onset temperature $T_{onset}$ of pairing in Bi 2212 in the *x-T* phase diagram derived from detailed ARPES measurements obtained from one crystal (Y. G. Zhong *et al*., [40]). Panel (c) shows the onset temperature $T^{ons}$ of pairing in YBCO inferred from intra-bilayer Josephson plasma detected by FIR spectroscopy (Dubroka *et al*., [38]).

Figure 8 displays 3 phase diagrams in the *x-T* or *p-T* plane (*x* or *p* is the hole density). Panel (a) compares the onset temperature of pairing $T_p$ (inferred from gap spectroscopy using STM [37]) vs the $T_c$ curve in the bilayer cuprate $Bi_2Ca_2SrCu_2O_8$ (Bi 2212). $T_p$ rises well above $T_c$ in the optimum and underdoped regimes. Panel (b) shows the onset temperature for pairing $T_{onset}$ vs. *x* derived from detailed ARPES measurements [40] (red symbols). Tuning of the hole density *x* over a broad range is achieved by *in situ* oxygen/vacuum annealing of a single crystal of Bi 2212. Within the error bars, the onset curve is closely similar to the curve derived from the Nernst effect (black circles). Panel (c) shows the onset temperature $T^{ons}$ of the intra-bilayer Josephson plasma detected by FIR spectroscopy in YBCO [38]. The plasma resonance, a collective Josephson mode of the pair condensate, requires phase rigidity between the two layers of each bilayer. Rigidity between adjacent bilayers appears only below $T_c$. Collectively, these results show that the pair condensate continues to exist well above $T_c$ but as a vortex liquid that lacks phase rigidity.

These considerations are important for interpretation of experiments in very high magnetic fields. While resistivity is by far the most commonly used probe in high fields, it lacks the ability to distinguish between the true normal state and the dissipative vortex liquid state when $H$ exceeds the melting field $H_m$. As such, an important unanswered question is whether high-field resistivity measurements in $YBa_2Cu_3O_y$ (YBCO) and Bi 2212 are probing the properties of the dissipative vortex liquid rather than the underlying normal state. By comparison, Nernst and magnetization experiments, which detect phase slips caused by vortex flow and diamagnetism, respectively, are more incisive probes for detecting the vortex liquid.

Figure 9 shows recent torque magnetization $M$ measured in fields up to 45 T at temperatures down to 0.5 K [41]. At each $T$, the sweep-up and sweep–down segments of $M$ (arrows) are hysteretic when the vortices are pinned (in a phase called the "vortex solid"). The vanishing of the hysteresis (at 45 T for $T = 0.5$ K) signals the melting of the vortex solid into a liquid. At the doping p = 0.11, the large negative value of M (diamagnetic response) implies that the vortex liquid survives well above 50 T. Interestingly, there are two vortex solid phases with distinct melting fields $H_{m1}$ and $H_{m2}$. This implies that in increasing $H$, the condensate undergoes an

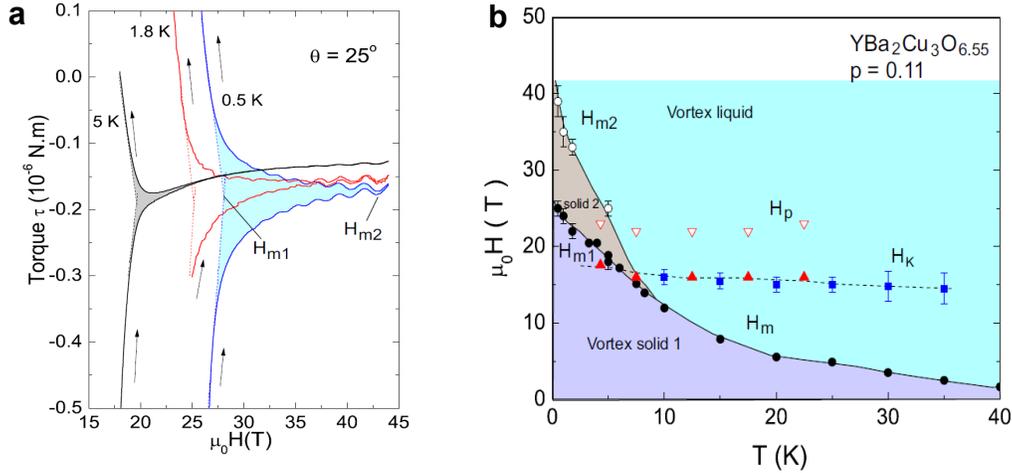

Figure 9 (a) Hysteretic torque magnetization curves measured on an untwinned YBCO crystal (p = 0.11) at 0.5, 1.8 and 5 K. At 0.5 K, the vortex solid 2 state persists to $H_{m2}$~ 45 T (light blue wedge). Above $H_{m2}$, the negative sign of the magnetization implies a vortex liquid that persists to much higher fields. Panel (b) shows the phase diagram in the $T$-$H$ plane for doping $p$ = 0.11. Below 10 K, two phases of the vortex solid with distinct melting fields are seen. The experiment also detects the crossover fields $H_K$ and $H_p$ related to the onset of charge ordering (but not to the pair condensate *per se*). The vortex liquid (light blue) extends to fields well above 45 T. (Fan Yu *et al*., [41]).

additional transition to accommodate the high field which leads to two distinct vortex solid phases whose exact nature remains unknown.

Significantly, quantum oscillations can be seen at fields as low as 27 T, which lies deep inside the vortex solid phase. This seems incompatible with the initial interpretation that the quantum oscillations occur in the normal state (above $H_{c2}$). Indeed, in the cuprate superconductors, the origin of quantum oscillations seems to be more subtle, interesting and, equally, the subject of renewed debate as new high field measurements are published. Torque experiments in higher fields (65 T and beyond) will help uncover the origin. Panel (b) displays the magnetic phase diagram inferred from the magnetization. Below 10 K, the melting field $H_m$ splits into the two branches [41]. The lower branch $H_{m1}$ terminates at 24 T which is sometimes erroneously identified with $H_{c2}$. The upper branch $H_{m2}$ increases rapidly to >42 T as $T\rightarrow 0$. The vortex liquid survives to fields well above 50 T. The data suggest that at $T = 0$, the vortex liquid exists over an extended field scale from 45 T to $H_{c2}$. Availability of DC fields of 65 T will allow this interesting phase of quantum matter to be investigated using a wider range of experimental probes.

Recent experiments have provided strong clues that may help unravel the puzzle of the pseudogap state. Previous results showed that in underdoped YBCO, *d*-wave superconductivity competes strongly with a weak, 2D charge density wave (CDW) state that appears at 150 K. The latter shows no sign of developing true long range order. Gerber et al. [42] carried out x-ray diffraction studies using a free-electron laser in a pulsed magnetic field H up to 28 Tesla to investigate CDW formation in $YBa_2Cu_3O_y$ with $y$ = 6.67. They report that when *H* exceeds 15 T, a 3D CDW emerges (at roughly 150 K) and develops long range order (Fig. 10). Despite the growth of the field-induced 3D CDW, superconducting correlations continue to exist up to 28 T consistent with a much higher $H_{c2}$ as implied in the phase diagram in Fig. 9b.

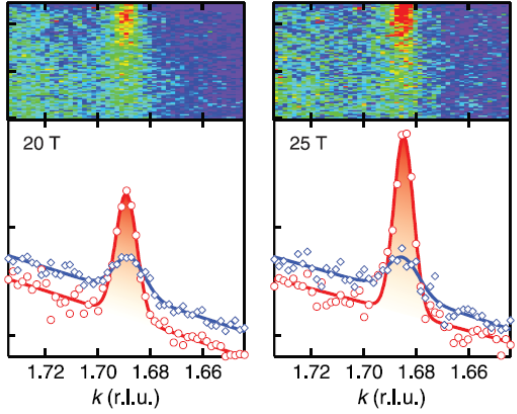

Figure 10: (A) CDW diffraction pattern (top panels) and the projected CDW peak profiles measured in YBCO at 10 K in a pulsed field of 20 and 25 T (Gerber et al.). The peaks are at $(hkl) = (0,k,-1)$ (red symbols) and $(0,k,-1/2)$ (blue) (Edkins *et al*. [43]).

A recent STM investigation of Bi 2212 in a magnetic field of 8.25 T (Edkins *et al*. [43]) has revealed charge modulations in the halo around a vortex core with wavevectors of $Q_P$ and $2 Q_P$. The authors argue that the field-induced CDW is actually a field-induced pair density wave (PDW) with 8 $CuO_2$ unit-cell periodicity with a predominantly *d*-symmetry form factor. These results imply that the ultimate state in cuprates in high fields is a PDW state involving Cooper pairing with finite momentum.

## 9. Pulsed or DC Field?

The community of high-field researchers was tasked with deciding whether resources should be directed towards building a 60 T DC magnet or a pulsed field magnet capable of reaching 150 Tesla. Participants who attended the high-field workshop "Exploring quantum phenomena and quantum matter in ultrahigh magnetic fields," held at NSF headquarters in Alexandria on Sep 21st and 22nd, 2017, discussed this issue extensively in two sessions that bracketed the invited talks. No obvious consensus emerged from the discussions. The reason for this ambivalence is that, judging from the past two decades, scientific progress and the process of serendipitous discovery in quantum matter have benefitted enormously from the synergistic exploitation of both pulsed fields and DC fields. The former provides a fleeting view of what lies in ultrahigh magnetic fields but its experimental purview is necessarily restricted by the short measurement time scales (<1 ms). To date, experimental probes in pulsed fields seem to be restricted to resistance, Hall effect, magnetization (both direct and torque measurements), magnetostriction, and optical spectroscopy (a few new scattering tools, e.g. neutron and x-ray scattering, are emerging). Apart from magnetization and magnetostriction, reliable thermodynamic measurements are largely precluded from pulsed field experiments. DC fields allow a much larger arsenal of probes to be brought to bear, with much better experimental resolution. Moreover, experiments can be performed at temperatures as low as 10 mK in a DC field (this is possibly its most crucial advantage). However, the restriction of DC magnets to lower maximum fields than can be achieved with pulsed magnets is an obvious restriction (Fig. 11).

Can a stand-alone pulsed field facility, decoupled from commensurate advances in DC field capabilities, be a transformative game changer in research on quantum phenomena? By its nature, the process of scientific discovery seems to require the nurturing of a broad range of expertise and technologies, all benefitting from findings achieved in allied areas and thriving by mutual competition. Funneling resources to enable a single capability to vault far ahead can land that capability or facility on an island in parameter space very remote from the main spheres of scientific interest. History provides an example. In the 1970's several labs worldwide focused considerable resources to build one-shot implosive magnets that achieved intense fields for 1-10 microseconds of several hundred teslas. Despite a decade of optical and resistance measurements, largely on bismuth, no discoveries or findings of lasting value in either fundamental physics or applied sciences have been documented. These pulsed field facilities had no discernible influence on the spate of discoveries in quantum matter that soon began in the early 1980s.

Recent progress in magnet technology at the NHMFL, both in Tallahassee and Los Alamos, suggest an interesting approach. The technology to construct both a 60 T DC magnet and a 130 T pulsed magnet are within reach. Both should be pursued in parallel to exploit the advantages and synergies existing between them.

In particular, strong progress is being made in DC superconducting magnets now that the high-temperature superconductors (HTS) based on REBCO (rare-earth barium copper oxide) are enabling magnets that can produce intense magnetic fields. The first HTS user magnet commissioned in Sendai, has now reached 24 T over 450 times. At the NHMFL, the maximum field has been raised by 33% to 32 T. It has reached its maximum field over 20 times. With funding from NSF, NHMFL has initiated the development of a 40-Tesla superconducting (SC) magnet. In addition, labs in France and China are also starting design of 40-Tesla SC magnets.

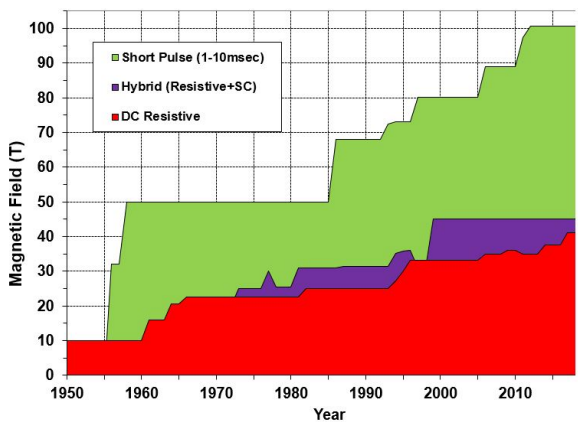

Figure 11: Peak magnetic fields available to researchers versus year of introduction for the three most utilized classes of magnets.

As for pulsed magnets, significant improvements in high-strength, high-conductivity materials and reinforcement materials were realized as part of the NHMFL's 100-T pulsed magnet completed in 2012. More recently, NHMFL has invested in increased in-house high-strength wire fabrication equipment (including a bull-block) and know-how which should enable the next generation of pulsed magnets with fields reaching 130 Tesla.

The fusion community is investing heavily in large-scale HTS magnet technology with multiple privately funded tokomaks being developed presently. An example is the SPARC design pursued jointly at MIT and Commonwealth Fusion Systems. The availability of HTS magnets with peak fields 10-13 Tesla will trigger a transformative advance in realizing commercial fusion energy. Current fusion designs (e.g. ITER) employ low-$T_c$ Nb$_3$Sn superconducting wires with peak fields of 5-6 Tesla. In a fusion device of dimension $R$, the energy gain goes as $R^{1.3}B^3$, and the power density varies as $RB^4$. See for e.g. the tutorial talk by Prof. Dennis Whyte [44]. Hence if the field $B$ can be increased 3-fold using REBCO-based magnets, it is theoretically possible to reduce the length scale $R$ (and the cost) by a factor of ~80, for the same power density. This promise has ignited strong interest in realizing compact fusion devices capable of producing 500 MW. Recently, the *Journal of Plasma Physics* devoted an entire issue to compact fusion reactors based on HTS magnets (vol. **86**, Issue 5, 2019). In turn the intense interest from the fusion community is driving further development of REBCO wires. This effort is resulting in magnet technology with multiple privately funded tokomaks being developed presently. This effort is resulting in dramatic increases in the amount of manufactured REBCO tape each year, which may lead to significant reductions in materials cost. The 40-Tesla projects and the large-volume magnet research from fusion labs should dovetail synergistically to enable the even higher field HTS magnet technology required for the middle part of a 60-Tesla resistive/superconducting hybrid magnet.

## Appendix A
In Appendix A, we explain some of the terms and underlying ideas encountered in the main text.

### A0: Magnetic field and Flux Quanta
A magnetic field *B* plays a central role in the experiments discussed because it is an inherently quantum probe of the electron fluid -- it causes the phase of the electron wave function to wind at a rate fixed by the strength of the field. At low temperatures and in high fields, the quantum granularity of a magnetic field becomes manifest and all-important. The magnetic flux (the product of *B* and a specified area) is actually comprised of a dense forest of irreducible objects called flux quanta $\phi_0 = h/e$ which are depicted as small vertical arrows (*h* is Planck's constant and *e* the electron charge). The discretization may be likened to the pixelation of a newsprint photograph under high magnification. The physical reality of the flux quantum first emerged in superconductivity. When a type II superconductor is exposed to a field, the flux entering the superconductor is physically discretized into a triangular array of vortices (supercurrent

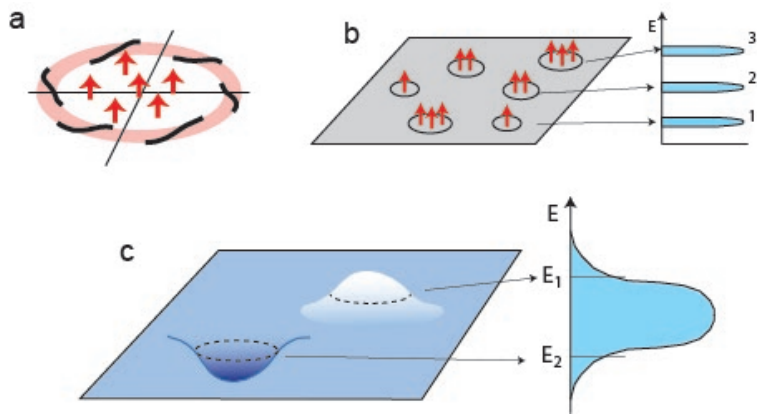

Figure A1 (a) A cyclotron orbit (pink hoop) encircling 6 flux quanta (red arrows). The phase of the wave-function winds 6 times around the orbit (black curve). (b) Grouping of orbits encircling 1, 2 or 3 flux quanta to define their respective Landau levels (bands shaded blue). (c) The golf-course topography in the presence of disorder. In the broadened Landau level (shaded blue), the energy of a contour (dashed curve) near a lake bottom ($E_2$) is lower than the average; around a hill top it is higher ($E_1$).

whirlpools in the quantum limit) each of which carries exactly one flux quantum (in a superconductor, $\phi_0 = h/2e$ ). The vortices have been imaged by decorating them with microscopic ferromagnetic particles ("nickel smoke") as well as by scanning tunneling microscopy (STM). In a normal metal, it is not possible to visualize the flux quanta. Nevertheless, the electrons can "count" them – flux quanta essentially define the allowed quantum states.

### A1: Landau levels and quantum oscillations
The formation of Landau levels is easiest to describe for electrons strictly confined to a two-dimensional sample (as in GaAs-AlGaAs quantum wells or graphene). In a magnetic field, electrons execute circular cyclotron orbits in the two-dimensional (2D) plane. In the classical (weak field) limit, the radii can assume any arbitrary value, but not in a strong magnetic field. A cyclotron orbit is allowed if, and only if, it encircles precisely an integral number *N* of flux quanta $\phi_0$. With $N\phi_0$ encircled, the phase of the electron's wavefunction winds by $2\pi N$ radians to match its starting value on completing the orbit (by contrast, non-integral numbers would lead to wave functions that are multivalued and ill-defined). Figure A1a shows a sketch of an orbit encircling 6 flux quanta. The phase (black curve) winds by $12\pi$ radians around the orbit. The subset of orbits encircling *N* quanta collectively define the $N^{th}$ Landau level (Fig. A1b). Adjacent Landau levels are separated in energy by a gap fixed by the cyclotron frequency with the *N*=0 level at the lowest energy. The number of electrons that can be accommodated in a given Landau level (its degeneracy) is equal to the number of flux quanta piercing the sample, i.e. each $\phi_0$ creates a quantum slot that can host one electron in each level. In a weak $B$, the degeneracy is low, so the electrons occupy many Landau levels (the highest level defines the chemical potential $\mu$). As $B$ increases, the degeneracy grows in proportion, causing the chemical potential, $\mu$, (the energy of the most energetic electrons) to drop to lower levels in a discontinuous way (jumps occur when the last electron leaves the highest Landau level). The sequential emptying of the highest levels leads to periodic oscillations in the observed resistance (or magnetization) when it is plotted versus $1/B$. Because the period is proportional to the cross-section area of the Fermi surface (FS), quantum oscillations have become the standard tool to measure the FS area in metals (see main text). In the limit of very strong fields, the lowest Landau level has enough states to accommodate all the electrons. This is called the quantum limit. Many of the experiments described are in this limit.

## A2: Quantum Hall Effect and Edge States

In the preceding Sec. A1, we assumed that each Landau level has negligible width (the 2D sample is perfectly uniform). In physical samples, the Coulomb potentials of nearby charged impurities expose the electrons to a smoothly undulating "golf course" topography featuring shallow minima (lakes) and maxima (hills) as sketched in Fig. A1c. In a strong magnetic field, electrons execute tight cyclotron orbits. The centers of the orbits drift in the direction **v** perpendicular to both **B** and the electric field **E** arising from the disorder potential (Fig. A2a). As the equipotential contours of the disorder potential are perpendicular to **E**, the orbit centers drift parallel to the contours. Within a Landau level, the orbits that trace contours around a lake lie lower in energy than the average, while contours encircling hill tops lie higher (Fig. A1c). This leads to broadening of the Landau levels and overlapping of states from adjacent levels. Now, in an increasing $B$, $\mu$ falls smoothly between adjacent levels (instead of executing abrupt jumps). During this smooth descent, the Hall resistivity displays the strictly quantized value $\rho_{xy} = h/e^2\nu$ where the integer $\nu$, "the filling factor", is the number of Landau levels whose average energy is less than $\mu$. In 2D, resistivity shares the same unit as resistance (Ohms). The combination $h/e^2 = 25.78$ k$\Omega$ is the quantum of resistance. This spectacular result -- the integer quantum Hall effect (IQHE) – is one of the major discoveries in the field of quantum matter. Its discovery was recognized by a Nobel Prize (awarded to von Klitzing). Today, it serves as the commercial standard for electrical resistance.

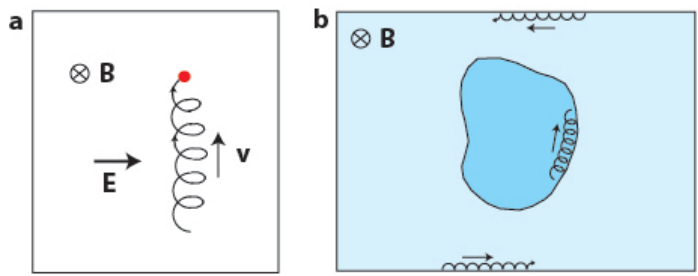

Figure A2. Panel (a): Classical motion of an electron (red dot) in a magnetic field **B** (directed into page) and an electric field **E** (to the right). The center of the cyclotron orbit drifts upwards with velocity **v**. Panel (b): In the 2D plane of a QHE sample, orbit centers hug the contours of a lake (dark blue region). At the boundaries, they trace skipping orbits (arrows).

Although the QHE is fundamentally quantum in character, the following discussion may help to introduce some of the concepts. It is instructive to view the broadened Landau level as a golf course partially submerged under water with $\mu$ defining the water level (all the lakes are connected underground so that the water level is the same everywhere). In the corresponding Landau level, all occupied states are under water. In the presence of the undulating topography, the water's edge traces out the contours corresponding to states at the chemical potential $\mu$. These states are the most important for carrying an injected electrical current. We assume that μ is initially in the lower half of the $(N+1)^{st}$ Landau level. As $B$ increases, the falling water level traces the low-energy contours encircling the lakes (Fig. A2b). Orbits hugging these contours cannot conduct an electrical current because the contours invariably form closed loops (we say the electrons are "localized"). As μ falls further to the upper half of the broadened $N^{th}$ level, the receding water level exposes the hill tops of its golf course. Again, orbits encircling the hill contours are localized. The absence of any orbit traversing the length of the 2D sample in the golf course (called the "bulk states") implies that the bulk is strictly insulating at low temperatures.

The novel feature in the QHE problem is the existence of conducting states at the edges of the sample. Termination of the 2D sample at its boundary is represented by a high wall surrounding the golf course. At the wall, the water's edge now defines new orbits that run the length of the sample known as chiral edge states. As shown in Fig. A2b, these edge states result from the repeated bouncing of the electron executing cyclotron motion against the hard wall potential (they are called skipping orbits by analogy with how a spinning flat stone skips on a lake). The velocity **v** of the orbit center is strictly unidirectional for fixed **B** (Fig. A2a), with the opposite direction forbidden ("chiral" here refers to the unidirectional propagation). The momentum of an electron in the edge mode cannot be reversed by collision with an impurity because the reversed-momentum state does not exist – the incident electron simply dances around the impurity to resume its original momentum. Because each completely filled Landau level gives rise to one chiral edge mode, and each mode yields the universal conductance $\sigma_0 = e^2/h$, the observed Hall resistivity has the quantized value $\rho_{xy} = e^2/h\nu$ when $\nu$ Landau levels are occupied. Throughout the plateau region, immunity from momentum-reversal scattering renders the material dissipationless (the observed diagonal resistivity $\rho_{xx}$ is rigorously zero). This implies that electrons move without "friction", or Joule heating is absent. Note that dissipationless electron transport is also observed in superconductors, but the physics is quite different in the superconducting state.

As $B$ continues to increase, μ eventually crosses the center of the $N^{th}$ Landau level. At the center, there exists at least one contour traced by the water's edge that neither encircles a hilltop nor a lake but runs across the bulk of the 2D sample. At this point, the bulk is able to conduct an electrical current, albeit along an interior path that is susceptible to momentum-reversal collisions, and hence dissipative (the resistivity $\rho_{xx}$ rises to a sharp peak while $\rho_{xy}$ loses its quantization). As μ sinks further into the lower half of the $N^{th}$ level, $\rho_{xx}$ again vanishes and $\rho_{xy}$ is quantized. The profile of $\rho_{xy}$ vs. $B$ is a rising staircase featuring broad quantized plateaus connected by nearly vertical step-rises which occur when μ crosses the center of a level. What happens when μ enters the lowest $N=0$ level is discussed next.

**A3: Fractional Quantum Hall and Composite Particles**

Discovery of the integer QHE was soon followed by the equally surprising discovery of the fractional quantum Hall effect (FQHE) in GaAs-AlGaAs quantum wells. When μ lies inside the lowest Landau level, Hall plateaus are observed whenever the filling factor assumes the odd-denominator fractions ($\nu = 1/3, 2/3, 1/5, 2/5, ...$). Unlike the states in the integer QHE, these fractional states arise from dominant Coulomb interaction between the electrons. The theory of the FQHE requires the full arsenal of techniques from quantum field theory. As an introduction, we discuss the Composite Particle approach to describe the $\nu = ½$ state and, in turn, to describe qualitatively how these strange new FQHE states originate.

When investigating a complicated many-body state, researchers focus solely on the small population of electronic excitations (quasiparticles) and ignore the vast sea of quiescent electrons because the applied external probes (injected current, photons, magnetic field, sound waves, etc) couple only to the excitations. The response of the excitations largely defines experimentally the nature of the many-body state. For example, the excitations in a familiar metal, say copper or gold, are strictly Landau quasiparticles which imitate real electrons in how they respond to electromagnetic fields, but are non-interacting unlike real electrons. All familiar properties of copper and gold devolve from the behavior of these non-interacting quasiparticles. In a superconductor, the excitations are the small population of broken electron pairs (Bogolyubov quasiparticles) in equilibrium with the collective pair condensate. In a thermal conductivity, ultrasonic absorption, or tunneling experiment, we detect the response of the Bogolyubov quasiparticles (the pair condensate ignores these probes).

At exactly ½ filling in the FQHE problem, there are 2 flux quanta for every electron in the sample. In the theory, the excitation is a composite particle formed by binding a physical electron to two flux quanta. (It is helpful to view each flux quantum as a vortex. The suppression of electron density in the cores of the vortices creates a deep potential well that traps an electron to form the composite particle.) Under exchange of two composite particles, their wave function acquires a minus sign consistent with fermion statistics. Because each composite fermion absorbs two flux quanta, no unbound flux quanta remain at $\nu = ½$. Hence composite fermions see zero (effective) magnetic field. The striking prediction is that when the applied field $B$ is incremented away from $\nu = ½$ filling, the extra unbound flux quanta inserted act as a reduced "effective field" $B_{eff}$ which cause the composite fermions to execute large cyclotron orbits. This has been observed in beautiful experiments in which the cyclotron orbits are perturbed by a weak periodic potential. More interestingly, in large $B_{eff}$, the composite fermions form their own Landau levels as in the IQHE and display the corresponding Hall plateaus in $\rho_{xy}$. At a higher meta-level, the sequence of FQHE states are theorized to be just the sequence of IQHE states generated by the emergent composite fermions. Indeed, a quantitative examination of $\rho_{xy}$ vs $B$ in the FQHE regime supports this point of view.

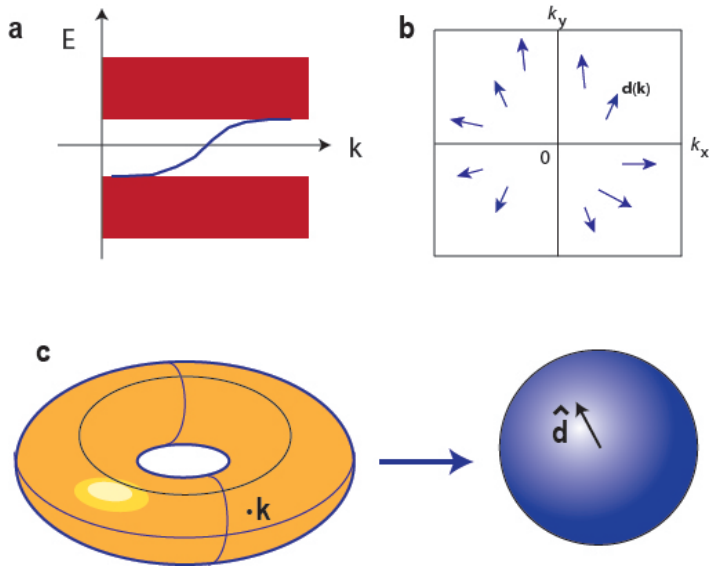

Figure A3. (a) A chiral edge mode (blue curve) traversing the energy gap separating two Landau level bands (deep red) in the $E$ vs $k$ plane. The edge mode disperses to the right (right-moving). (b) The 3D vector field $\mathbf{d}(\mathbf{k})$ defined by $H$ in the 2D ($k_x$, $k_y$) plane (BZ). States on the right edge are identified with states on the left (same for top and bottom edges), so the edges can be glued together to form a torus. (c) The Gauss map from the BZ torus (orange) to the unit-radius sphere (purple). Each point $\mathbf{k}$ on the BZ torus is mapped by the Hamiltonian to a unit vector $\hat{\mathbf{d}}$ on the sphere. The Chern number $\mathcal{C}$ is the number of times the map wraps around the sphere as $\mathbf{k}$ sweeps through all the states on the torus.

## A4: Chern number, Chern Insulator

In a strong magnetic field, a Landau level (in a 2D sample) may be regarded as a single band of ("bulk") states separated from adjacent levels by large gaps set by the cyclotron frequency (Fig. A3a). Conventionally, there are no conducting states traversing these gaps (hence when µ lies inside the gap, the sample is an insulator at low temperature). However, chiral edge modes exist at the boundary (see Sec. A2). In a plot of energy vs. momentum ($E$ vs. $\mathbf{k}$), a chiral mode appears as a conducting state that traverses the gap (blue curve in Fig. A3a). For fixed $\mathbf{B}$, the mode disperses in only one direction (right-moving, say) explicitly breaking time-reversal invariance. This is the archetypal example of a Chern insulator – the bulk is insulating but the sample displays a quantized Hall conductance because of the chiral edge modes. The momenta of the electronic states in a 2D material are represented as vectors **k** in the 2D plane in momentum space ($k_x$, $k_y$) called the Brillouin zone (BZ). Although drawn as a square, the BZ actually lives on the surface of a torus (orange inner tube in Fig. A3c) because states that move beyond the right boundary of the BZ reappear on the left boundary like the ball in the video game *Pong* (likewise for the top and bottom boundaries). The electronic states are calculated from the Hamiltonian *H*, which is a $2 \times 2$ matrix.

Mathematically, all $2 \times 2$ matrices can be expressed as an expansion in terms of the standard Pauli matrices $\boldsymbol{\sigma} = (\sigma_1, \sigma_2, \sigma_3)$ plus the unit matrix **1** which we ignore. Hence the Hamiltonian has the form $H = \mathbf{d}(\mathbf{k}) \cdot \boldsymbol{\sigma}$. The vector $\mathbf{d} = (d_1, d_2, d_3)$ is specific to the material of interest. In the 2D plane ($k_x$, $k_y$), $\mathbf{d}(\mathbf{k})$ defines a field of 3D vectors (Fig. A3b). Next we regard $\mathbf{d}(\mathbf{k})$ as a (Gauss) map from the BZ torus to a unit-radius sphere defined by its unit vector $\hat{\mathbf{d}}(\mathbf{k}) = \mathbf{d}/|\mathbf{d}|$. The Hamiltonian maps each point **k** on the torus to a point $\hat{\mathbf{d}}(\mathbf{k})$ on the sphere (Fig. A3c). As **k** wanders over the BZ torus, $\hat{\mathbf{d}}(\mathbf{k})$ varies smoothly over the unit sphere. The pattern of how $\hat{\mathbf{d}}(\mathbf{k})$ varies tells us whether the material is topologically trivial or non-trivial.

Imagine initiating a dab of paint somewhere on the BZ torus and allowing it to expand until the whole torus is covered. Under the map, the expansion generates an image patch that spreads on the unit sphere where $\hat{\mathbf{d}}(\mathbf{k})$ lives. If the image patch ends up covering the entire sphere, the material is topologically non-trivial. By contrast, in a topologically trivial material, the image patch initially grows, but subsequently retreats to the starting point, erasing all traces of the initial coverage. The number of times the map fully wraps around the sphere is known as the Chern number $\mathcal{C}$ (or winding number). $\mathcal{C}$ equals 1 in the previous case and zero in the latter. Its topological nature is clear. Just as a knot in a closed loop cannot be removed without cutting the string, here we cannot change the winding number without closing the original gap. We describe how $\mathcal{C}$ determines the 2D Hall effect after introducing the Berry curvature below.

**A5: Berry curvature and Weyl nodes**

To specify the position **r** of an electron in a periodic lattice, it would appear necessary to state both the unit cell $\mathbf{R_n}$ (the Wannier coordinate) and the intracell coordinate **X** that locates the electron within the cell (i.e. $\mathbf{r} = \mathbf{R_n} + \mathbf{X}$). For 4 or more decades, **X** was completely discarded (with the exception of the celebrated Luttinger-Karplus theory). Its importance only emerged in the late 1990s after ramifications of the Berry phase in crystals became better understood. The coordinate **X**, relabeled as **A(k)**, defines a vector potential that lives strictly in **k** space. Its derivative with respect to ($k_x$, $k_y$, $k_z$) (the curl of **A(k)**) defines a vector field $\boldsymbol{\Omega}(\mathbf{k})$ called the Berry curvature. As the curl operation suggests, $\boldsymbol{\Omega}(\mathbf{k})$ acts on a wave packet just like an effective magnetic field (again living in **k** space). It imparts a phase shift (the Berry phase) to a wave packet accelerating in **k** space. In an applied electric field **E**, the cross-product of **E** and $\boldsymbol{\Omega}(\mathbf{k})$ leads to an anomalous (Luttinger) velocity that bends the orbit away from the direction of acceleration to yield an anomalous Hall effect.

In principle, the Berry curvature is present in *all* crystalline materials. However, if the material respects both time-reversal invariance and inversion symmetry, $\boldsymbol{\Omega}(\mathbf{k})$ has the value zero at all **k**. Once we break either symmetry (notably time-reversal invariance by applying **B**), $\boldsymbol{\Omega}(\mathbf{k})$ emerges to produce effects on the electronic properties equivalent to physical magnetic fields as large as 300 Tesla.

Returning to the discussion of Landau bands (Fig. A3a), the Berry curvature is finite everywhere in the BZ because **B** breaks time-reversal invariance. In analogy with ordinary flux, we form the product of $\Omega_z$ and a specified area in **k** space to define the Chern flux. Integrated over the whole

BZ, the Chern flux yields an integer (0, 1, 2,…) that is just the Chern number $\mathcal{C}$. (Because of the torus topology, an attempt to define **A(k)** valid over the whole torus actually encounters a conundrum called "obstruction", just as in the Dirac monopole problem. As in that problem, the resolution leads to the quantization of $\mathcal{C}$.) In a seminal paper, Thouless and collaborators showed that the 2D Hall conductivity is directly related to $\mathcal{C}$ by the equation $\sigma_{xy} = \mathcal{C}e^2/h$. This key result implies that $\mathcal{C}$ counts the number of edge modes intersected by µ.

In a 3D Dirac semimetal, the breaking of either time-reversal invariance or inversion symmetry splits each Dirac cone into two Weyl cones with opposite "handedness" or chirality (1 and -1). One Weyl node acts like a monopole source for Berry curvature in **k** space while the other like a sink. The most prominent feature of the Weyl nodes appears when a strong magnetic field **B** leads to Landau level formation. Along the direction of **B**, the lowest Landau levels are strictly unidirectional ("chiral" like the edge states in the QHE, but here pertaining to the 3D bulk states). In the lowest Landau level, the right-handed Weyl fermions (with chirality +1) have a velocity strictly parallel to **B** whereas the left-handed ones (chirality -1) have the opposite velocity. This realizes a universe in which we have equal populations of one-dimensional massless fermions either right-moving or left-moving (with respect to **B**), which is the setting for the chiral anomaly. When **E** is applied parallel to **B**, the right-moving population, say, increases at the expense of the left-movers. This produces a new current called the axial current.

For more detailed and technical treatments of these topics, we refer the reader to
1. Horst L. Stormer, "Nobel Lecture: The fractional quantum Hall effect," *Rev. Mod. Phys.* **71**, 875 (1999).
2. Zyun F. Ezawa, *Quantum Hall Effects* (World Scientific Press, 2008).
3. Jainendra K. Jain, *Composite Fermions* (Cambridge Univ. Press, 2012).
4. Xiao-Liang Qi, Taylor L. Hughes, and Shou-Cheng Zhang, "Topological field theory of time-reversal invariant insulators," *Phys. Rev. B* **78**, 195424 (2008).
5. B. Andrei Bernevig and Taylor L. Hughes, *Topological Insulators and Topological Superconductors* (Princeton University Press, 2013).
6. D. Xiao, M. C. Chang, and Q. Niu, "Berry phase effects on electronic properties," *Rev. Mod. Phys.* **82**, 1959 (2010).
7. Naoto Nagaosa *et al.*, Anomalous Hall Effect, *Rev. Mod. Phys.* **82**, 1539 (2010).

**Appendix B**
Participants in the workshop *Exploring quantum phenomena and quantum matter in ultrahigh magnetic fields*, National Science Foundation, Alexandria, Sep. 21--22 (2017)

Analytis, James  (Univ. Calif. Berkeley)
Aronson, Megan (Texas A&M)
Bernevig, Andrei (Princeton)
Bird, Mark (Florida State, NHMFL)
Boebinger, Greg (Florida State, NHMFL)
Broholm, Collin (Johns Hopkins)
Dean, Cory (Columbia)
Analytis, James  (UC Berkeley)
Aronson, Megan (Texas A&M)
Bernevig, Andrei (Princeton)
Bird, Mark (Florida State, NHMFL)
Boebinger, Greg (Florida State, NHMFL)
Broholm, Collin (Johns Hopkins)
Dean, Cory (Columbia)
Freedman, Danna (Northwestern)
Fu, Liang (MIT)
Greene, Laura (Florida State, NHMFL)
Gupta, Ramesh (Brookhaven Nat. Lab.)
Hwang, Harold (Stanford)
Kim, Philip (Harvard)
Lee, Minhyea (U. Colorado, Boulder)
Li, Lu (U Mich, Ann Arbor)
Michael, Philip (MIT)
Mielke, Charles (Los Alamos Nat. Lab.)
Minervini, Joseph (MIT)
Musfeldt, Janice (U. Tennessee)
Oh, Sean (Rutgers)
Ong, N. Phuan (Princeton)
Ramshaw, Brad (Cornell)
Sachdev, Subir (Harvard)
Sebastian, Suchitra (Cambridge Univ)
Trivedi, Nandini (Ohio State)
Tyson, Trevor (NJ Inst. Tech.)
Vishwanath, Ashvin (Harvard)
Young, Andrea (UC Santa Barbara)
Zapf, Vivien (Los Alamos Nat. Lab.)

**Appendix C**:
Schedule of Talks

## Exploring quantum phenomena and quantum matter in ultrahigh magnetic fields
Alexandria, Sep 21-22, 2017

## Thursday, Sep. 21st, 2017

8:45 -- 9:00                 *Welcome and Introduction*

**1. Graphene Physics (Chair: Collin Broholm)**

| Time | Speaker | Title |
|---|---|---|
| 9:00 - 9:30 | Cory Dean | Correlated electronic states in high magnetic fields |
| 9:30 - 10:00 | Andrea Young | Fractional Chern insulators in van der Waals heterostructures |
| 10:00 - 10:30 | Philip Kim | van der Waal heterostructures at the extreme quantum limit |

10:30 - 10:50    *Break*

**2. Quantum and Topological Materials 1 (Chair: Brad Ramshaw)**

| Time | Speaker | Title |
|---|---|---|
| 10:50 - 11.20 | James Analytis | Intertwined and topological quantum states in high magnetic fields |
| 11:20 - 11:50 | Danna Freedman | Studies of Inorganic Materials at Extreme Conditions |

12:00 - 1:30    *Lunch*

**3. Superconductivity (Chair: Lu Li)**

| Time | Speaker | Title |
|---|---|---|
| 1:30 -- 2:00 | Subir Sachdev | Topology and criticality in the high temperature superconductors |
| 2:00 - 2:30 | Brad Ramshaw | Crossing the dome: evolution of the Fermi surface across optimal doping in high-Tc superconductors |
| 2:30 - 3:00 | Suchitra Sebastian | Unmasking unconventional phenomena in strongly correlated materials using high magnetic fields |

3:00 - 3:20    *Break*

**4. Prospects in high magnetic fields (Chair: Harold Hwang)**

| Time | Speaker | Title |
|---|---|---|
| 3:20 - 3:50 | Mark Bird | 60 T dc: harnessing the power of high temperature superconductors |
| 3:50 - 4:20 | Chuck Mielke | 135T pulsed: harnessing state-of-the-art nanocomposites and unique MagLab infrastructure. |
| 4:20 - 4:50 | Greg Boebinger | World Leadership and the Complementarity of a 60T DC Magnet and a 135T Pulsed Magnet |

4:50 - 5:40    **5 Panel Discussion 1 (Chair: N. Phuan Ong)**
Bird, Boebinger, Gupta, Michael, Mielke

## Friday, Sept. 22nd, 2017

**6. Quantum and Topological Materials 2 (Chair: Andrea Young)**

| | | |
|---|---|---|
| 9:00 -- 9:30 | Ashvin Vishwanath | Emergent Dirac Fermions |
| 9:30 - 10:00 | Liang Fu | Giant thermoelectric effect in Dirac materials under a quantizing magnetic field |
| 10:00 - 10:30 | Andre Bernevig | Topological Quantum Chemistry |

10:30 - 10:50   *Break*

**7. Synthetic structures (Chair: Minhya Lee )**

| | | |
|---|---|---|
| 10:50 - 11:20 | Harold Hwang | Complex oxide heterostructures |
| 11:20 - 11:50 | Sean Seongshik Oh | Topological quantum effects in thin film topological insulators at the magnetic quantum limit |

12:00 -- 1:30   *Lunch*

**8. Quantum and Topological Materials 3 (Chair: Meigan Aronson)**

| | | |
|---|---|---|
| 1:30 - 2:00 | Janice Musfeldt | Multiferroics in high magnetic fields |
| 2:00 - 2:30 | Nandini Trivedi | Magnetism and Superconductivity in Chern Insulators |
| 2:30 - 3:00 | Vivien Zapf | Overview of Quantum Materials Research at the NHMFL |

3:00 - 3:20   *Break*

**9. Quantum and Topological Materials 4 (Chair: Sean Oh)**

| | | |
|---|---|---|
| 3:20 - 3:50 | Collin Broholm | New opportunities to form and probe quantum matter through ultrahigh field neutron scattering |
| 3:50 - 4:20 | Minhyea Lee | Anomalous thermal conductivity in the honeycomb magnet $RuCl_3$ |
| 4:20 - 4:50 | Lu Li | Correlated Topological Materials in High Magnetic Fields |

**10. Panel discussion 2 (Chair: N. Phuan Ong)**

4:50 - 5:40   Young, Sebastian, Li, Zapf, Kim

# References

1. Liang Fu, "Topological crystalline insulators," *Phys. Rev. Lett*. **106**, 106802 (2011).
2. T.H. Hsieh, H. Lin, J. Liu, W. Duan, A. Bansil, and L. Fu, "Topological crystalline insulators in the SnTe material class." *Nat. Commun*. **3**, 982 (2012).
3. Y. Tanaka, *et al*. "Experimental realization of a topological crystalline insulator in SnTe." *Nat. Phys*. **8**, 800 (2012).
4. P. Dziawa, *et al*. "Topological crystalline insulator states in $Pb_{1-x}Sn_xSe$." *Nat. Mater*. **11**, 1023–1027 (2012).
5. Tian Liang, Quinn Gibson, Jun Xiong, Max Hirschberger, Sunanda P. Koduvayur, R. J. Cava and N. P. Ong, "Evidence for massive bulk Dirac Fermions in $Pb_{1-x}Sn_xSe$ from Nernst and thermopower experiments," *Nature Commun*. **4**:2696  doi: 10.1038/ncomms3696 (2013).
6. B. Skinner and L. Fu, "Large, nonsaturating thermopower in a quantizing magnetic field," *Science Advances* **4**, eaat2621 (2018).
7. G. Li, Z. Xiang, F. Yu, T. Asaba, B. Lawson, P. Cai, C. Tinsman, A. Berkley, S. Wolgast, Y. S. Eo, D. J. Kim, C. Kurdak, J. W. Allen, K. Sun, X. H. Chen, Y. Y. Wang, Z. Fisk, and L. Li, "Two-dimensional Fermi surfaces in Kondo insulator $SmB_6$," *Science* **346**, 1208 (2014).
8. B. S. Tan, Y.-T. Hsu, B. Zeng, M. Ciomaga Hatnean, N. Harrison, Z. Zhu, M. Hartstein, M. Kiourlappou, A. Srivastava, M. D. Johannes, T. P. Murphy, J.-H. Park, L. Balicas, G. G. Lonzarich, G. Balakrishnan, Suchitra E. Sebastian, "Unconventional Fermi surface in an insulating state," *Science* **349**, 287 (2015).
9. Z. Xiang, B. Lawson, T. Asaba, C. Tinsman, Lu Chen, C. Shang, XH Chen, and Lu Li, "Bulk Rotational Symmetry Breaking in Kondo Insulator $SmB_6$," *Phys. Rev. X* **7**, 031054 (2017).
10. Z. Xiang, Y. Kasahara, B. Lawson, T. Asaba, C. Tinsman, Lu Chen, K. Sugimoto, H. Kawaguchi, Y. Sato, G. Li, S. Yao, Y. L. Chen, F. Iga, Y. Matsuda, and Lu Li, "Quantum Oscillations of Electrical Resistivity in an Insulator," *in press, Science.*
11. J. Xiong, S. K. Kushwaha, T. Liang, J. W. Krizan, M. Hirschberger, W. Wang, R. J. Cava, and N. P. Ong, "Evidence for the chiral anomaly in the Dirac semimetal $Na_3Bi$," *Science* **350**, 413 (2015).
12. Q. Li, D. E. Kharzeev, C. Zhang, Y. Huang, I. Pletikosić, A. V. Fedorov, R. D. Zhong, J. A. Schneeloch, G. D. Gu, and T. Valla, "Chiral Magnetic Effect in $ZrTe_5$," *Nat. Phys*. **12**, 550 (2016).
13. Tian Liang, Jingjing Lin, Quinn Gibson, Satya Kushwaha, Minhao Liu, Wudi Wang, Hongyu Xiong, Jonathan A. Sobota, Makoto Hashimoto, Patrick S. Kirchmann, Zhi-Xun Shen, R. J. Cava, and N. P. Ong, "Anomalous Hall Effect in $ZrTe_5$,'' *Nature Physics* **14**, 451 (2018).
14. M. Hirschberger, S. Kushwaha, Z. Wang, Q. Gibson, S.Liang, C. A. Belvin, B. A. Bernevig, R. J. Cava, and N. P. Ong, "The Chiral Anomaly and Thermopower of Weyl Fermions in the Half-Heusler GdPtBi," *Nat. Mater*. **15**, 1161 (2016).
15. X. Huang, L. Zhao, Y. Long, P.Wang, D. Chen, Z. Yang, H. Liang, M. Xue, H. Weng, Z. Fang, X. Dai, and G. Chen, "Observation of the Chiral-Anomaly-Induced Negative Magnetoresistance in 3D Weyl Semimetal TaAs," *Phys. Rev. X* **5**, 031023 (2015).
16. C.-L. Zhang et al., "Signatures of the Adler-Bell-Jackiw Chiral Anomaly in a Weyl Fermion Semimetal," *Nat. Commun*. **7**, 10735 (2016).
17. R. D. dos Reis, M. O. Ajeesh, N. Kumar, F. Arnold, C. Shekhar, M. Naumann, M. Schmidt, M. Nicklas, and E. Hassinger, "On the Search for the Chiral Anomaly in Weyl Semimetals: The Negative Longitudinal Magnetoresistance," *New J. Phys*. **18**, 085006 (2016).

18. Sihang Liang, Jingjing Lin, Satya Kushwaha, Jie Xing, Ni Ni, R. J. Cava, and N. P. Ong, "Experimental Tests of the Chiral Anomaly Magnetoresistance in the Dirac-Weyl Semimetals $Na_3Bi$ and GdPtBi," *Phys. Rev.* X **8**, 031002 (2018).
19. B. Z. Spivak and A. V. Andreev, "Magnetotransport phenomena related to the chiral anomaly in Weyl semimetals," *Phys. Rev.* B **93**, 085107 (2016).
20. S.V. Syzranov, Ya.I. Rodionov and B. Skinner, "Adiabatic dechiralisation and thermodynamics of Weyl semimetals," cond-mat arXiv:1807.03778v1.
21. K. S. Novoselov et al., *Science* **306**, 666 (2004).
22. Y. Zhang, J. Tan, H. L. Stormer, and P. Kim, *Nature* **438**, 201 (2005).
23. L. A. Ponomarenko, R. V. Gorbachev, G. L. Yu, D. C. Elias, R. Jalil, A. A. Patel, A. Mishchenko, A. S. Mayorov, C. R.Woods, J. R.Wallbank, M. Mucha-Kruczynski, B. A. Piot, M. Potemski, I. V. Grigorieva, K. S. Novoselov, F. Guinea, V. I. Fal'ko and A. K. Geim, "Cloning of Dirac fermions in graphene superlattices," *Nature* **497**, 594 (2013).
24. C. R. Dean, L. Wang, P. Maher, C. Forsythe, F. Ghahari, Y. Gao, J. Katoch, M. Ishigami, P. Moon, M. Koshino, T. Taniguchi, K.Watanabe, K. L. Shepard, J.Hone and P. Kim "Hofstadter's Butterfly and the Fractal Quantum Hall Effect in Moire Superlattices," *Nature* **497**, 598 (2013).
25. B. Hunt, J. D. Sanchez-Yamagishi, A. F. Young, M. Yankowitz, B. J. LeRoy, K. Watanabe, T. Taniguchi, P. Moon, M. Koshino, P. Jarillo-Herrero, and R. C. Ashoori, "Massive Dirac Fermions and Hofstadter Butterfly in a van der Waals Heterostructure," *Science* **340**, 1427 (2013).
26. E. M. Spanton, A. A. Zibrov, H. Zhou, T. Taniguchi, K. Watanabe, M. P. Zaletel, and A. F. Young, "Observation of fractional Chern insulators in a van der Waals heterostructure," *Science,* **360**, 62 (2018).
27. Lei Wang, Yuanda Gao, Bo Wen, Zheng Han, Takashi Taniguchi, Kenji Watanabe, Mikito Koshino, James Hone, Cory R. Dean, "Evidence for a fractional fractal quantum Hall effect in graphene superlattices," *Science* **350**, 1231 (2015).
28. S. E. Sebastian, N. Harrison, C. D. Batista, L. Balicas, M. Jaime, P. A. Sharma, N. Kawashima, and I. R. Fisher, "Dimensional reduction at a quantum critical point," *Nature* **441**, 617 (2006).
29. S. Haravifard, *et al*. "Crystallization of spin superlattices with pressure and field in the layered magnet $SrCu_2(BO_3)_2$," *Nat. Commun.* **7**:11956 doi: 10.1038/ncomms11956 (2016).
30. J. Romhányi, K. Penc, and R. Ganesh, "Hall effect of triplons in a dimerized quantum magnet," *Nat. Commun.* **6**:6805 doi: 10.1038/ncomms7805 (2015).
31. Jiun-Haw Chu, James G Analytis, Kristiaan De Greve, Peter L McMahon, Zahirul Islam, Yoshihisa Yamamoto, Ian R Fisher. "In-plane resistivity anisotropy in an underdoped iron arsenide superconductor" *Science* **329**, 824-826 (2010).
32. F. Ronning, T. Helm, K. R. Shirer, M. D. Bachmann, L. Balicas, M. K. Chan, B. J. Ramshaw, R. D. McDonald, F. F. Balakirev, M. Jaime, E. D. Bauer, and P. J. W. Moll, "Electronic in-plane symmetry breaking at field-tuned quantum criticality in $CeRhIn_5$," *Nature* **548**, 313 (2017).
33. P. A. Lee, N. Nagaosa, and X.G. Wen, "Doping a Mott Insulator: Physics of high-temperature superconductivity," *Rev. Mod. Phys.* **78**, 17 (2006).
34. Eduardo Fradkin, Steve A. Kivelson, and John M. Tranquada, "Colloquium: Theory of intertwined orders in high temperature superconductors," *Rev. Mod. Phys.* **87**, 457 (2015).

35. Yayu Wang, Lu Li and N. P. Ong, "The Nernst effect in high-$T_c$ superconductors", Phys. Rev. B **73**, 024510 (2006).
36. Lu Li, Yayu Wang, Seiki Komiya, Shimpei Ono, Yoichi Ando, G. D. Gu and N. P. Ong, "Diamagnetism and Cooper pairing above $T_c$ in cuprates", *Phys. Rev.* B **81**, 054510 (2010).
37. K. K. Gomes, A. N. Pasupathy, A. Pushp, S. Ono, Y. Ando, and A. Yazdani, "Visualizing pair formation on the atomic scale in the high-Tc superconductor $Bi_2Sr_2CaCu_2O_{8+\delta}$," *Nature* **447**, 569 (2007).
38. A. Dubroka, M. Rössle, K. W. Kim, V. K. Malik, D. Munzar, D. N. Basov, A. A. Schafgans, S. J. Moon, C. T. Lin, D. Haug, V. Hinkov, B. Keimer, T. Wolf, J. G. Storey, J. L. Tallon, and C. Bernhard, "Evidence of a Precursor Superconducting Phase at Temperatures as High as 180 K in $R Ba_2Cu_3O_{7-\delta}$ ($R$ = Y, Gd, Eu) Superconducting Crystals from Infrared Spectroscopy," *Phys. Rev. Lett.,* **106**, 047006 (2011).
39. Bernhard Keimer, reported at the *Symposium on Emergent Quantum Materials*, University of Tokyo, Jan. 18-20, 2017, *unpublished*.
40. Y. Zhong, J. Guan, X. Shi, J. Zhao, Z. Rao, C. Tang, H. Jiu, G. D. Gu, Z. Weng, Z. Wang, T. Qian, Y. Sun, and H. Ding, "Continuous doping of a cuprate surface: new insights from in-situ ARPES," *preprint,* 2018.
41. F. Yu, M. Hirschberger, T. Loew, G. Li, B. J. Lawson, T. Asaba, J. B. Kemper, T. Liang, J. Porras, G. S. Boebinger, J. Singleton, B. Keimer, L. Li, and N. P. Ong, "Magnetic phase diagram of underdoped $YBa_2Cu_3O_y$ inferred from torque magnetization and thermal conductivity," *Proc. Natl. Acad. of Sci.* **113**, 12667 (2016).
42. S. Gerber, H. Jang, H. Nojiri, S. Matsuzawa, H. Yasumura, D. A. Bonn, R. Liang, W. N. Hardy, Z. Islam, A. Mehta, S. Song, M. Sikorski, D. Stefanescu, Y. Feng, S. A. Kivelson, T. P. Devereaux, Z. X. Shen, C. C. Kao, W. S. Lee, D. Zhu, and J. S. Lee, "Three-dimensional charge density wave order in $YBa_2Cu_3O_{6.67}$ at high magnetic fields," *Science* **350**, 949 (2015).
43. S.D. Edkins, A. Kostin, K. Fujita, A. P. Mackenzie, H. Eisaki, S. Uchida, M. J. Lawler, E-A. Kim, J.C. Séamus Davis and M. H. Hamidian, "Magnetic-field Induced Pair Density Wave State in the Cuprate Vortex Halo," *Science* **364***,* 976-980 (2019).
    *DOI: 10.1126/science.aat1773*
44. D. Whyte, *Breakthrough in Nuclear Fission?*
    https://www.youtube.com/watch?v=KkpqA8yG9T4&t=4711s

**Acknowledgment of support**
We acknowledge a grant from the U.S. National Science Foundation (DMR-1745525) which supported the workshop "*Exploring quantum phenomena and quantum matter in ultrahigh magnetic fields*," held at Alexandria, Sep. 21, 22 (2017), as well as the preparation of this report. N.P.O. acknowledges support from the Gordon and Betty Moore Foundation's EPiQS Initiative (Grant GBMF4539). L.L. acknowledges support from a grant from NSF (DMR-1707620).